\documentclass[12pt]{article}
\usepackage{epsfig}
\usepackage{graphicx}
\usepackage{amsmath, amsfonts}
\usepackage{amssymb}
\usepackage{latexsym}
\usepackage{color}
\usepackage{mathrsfs}
\usepackage{natbib}
\usepackage{hyperref}
\usepackage{appendix}
\usepackage{dsfont}
\usepackage[left=3cm,top=4cm,right=3cm,bottom=3cm]{geometry}
\RequirePackage[caption=false]{subfig}
\usepackage{lscape}
\usepackage{dsfont}
\usepackage{rotating}

\newcommand{\de}{\mathrm{d}}
\DeclareMathOperator{\tf}{\mathfrak{t}}

\newcommand{\LL}{\mathds{L}}
\newcommand{\F}{\mathds{F}}
\newcommand{\T}{\mathcal{T}}
\newcommand{\ee}{\mathds{E}}

\newcommand{\R}{\mathds R}

\DeclareMathOperator{\cov}{\mathds{C}ov}

\begin{document}
\title{\LARGE \bf 
Function-valued marked spatial point processes on linear networks: application to urban cycling profiles}
\date{April 29, 2024}
\maketitle
\begin{center}
{{\bf Matthias Eckardt$^{a}$},  {\bf Jorge Mateu$^{b}$}and {\bf Mehdi Moradi$^{c}$}\\
\noindent $^{\text{a}}$ Chair of Statistics, Humboldt-Universit\"{a}t zu Berlin, Berlin, Germany\\
\noindent $^{\text{b}}$ Department of Mathematics}, University Jaume I, Castell\'{o}n, Spain\\
\noindent $^{\text{c}}$  Department of Mathematics and Mathematical Statistics, Ume{\r a} University, Ume{\r a}, Sweden
\end{center}
%{\color{red} }

\abstract[Summary]{
In the literature on spatial point processes, there is an emerging challenge in studying marked point processes with points being labelled by functions. In this paper, we focus on point processes living on linear networks and, from distinct points of view, propose several marked summary characteristics that are of great use in studying the average association and dispersion of the function-valued marks. Through a simulation study, we evaluate the performance of our proposed marked summary characteristics, both when marks are independent and when some sort of spatial dependence is evident among them. Finally, we employ our proposed mark summary characteristics to study the spatial structure of urban cycling profiles in Vancouver, Canada.
}


\section{Introduction}\label{sec:intro}

Massive data collected from wearable sensors or automated tracking devices on various topics has become omnipresent in the last few years. Including human or animal movement tracks and real-time measurements of physiological and health-related parameters as typical examples, any such data is commonly characterised in its simplest case by a continuously recorded outcome collected over time.  In most cases, the exact timestamp is often also enriched by the geographical location of the object under study at that time, such as in the case of data derived from movement tracking of wildlife animals. In consequence, the statistical object of interest is often not only to understand the variation in the observed track profiles but also to investigate their spatial interrelations. 
One commonly applied statistical framework to analyse tracking data is to formalise the outcome as a realisation of a function-valued quantity and to apply different functional (shape) data analysis methods  \citep[see][]{DANNENMAIER2020182, doi:10.1080/09291016.2021.1929673,10.1111/biom.13706}. Apart from the non-spatial classification and modelling of functional data through multivariate functional data analysis methods and functional regression approaches, there have been some developments in spatial methods which allow for the analysis of the spatial dependence and variation of the observed function-valued {outcomes}. Combining the ideas from classic spatial data analysis \citep{cressie93} and functional data analysis \citep{RamsaySilverman2005}, such functional spatial data \citep{https://doi.org/10.1002/env.1003} approaches aim to investigate and quantify the interrelation between the spatially explicit function-valued outcomes by taking the spatial closeness between the objects under study into account. {Dominated by methods} for functional geostatistical and functional {areal data,} including functional kriging  \citep{doi:https://doi.org/10.1002/9781119387916.ch13, doi:https://doi.org/10.1002/9781119387916.ch3, doi:https://doi.org/10.1002/9781119387916.ch4},  functional mantel tests \citep{giraldo2018mantel}, and functional CAR \citep{doi:10.1080/01621459.2015.1042581} or SAR \citep{fsar} models, only a small (but currently growing) body of the literature {has covered} methods for spatial point processes in which each point location is augmented by at least one function-valued quantity, commonly denoted as mark \citep{Eckardt2023MultiFunctionMarks, Ghorbani2020, Comas2013, Comas2011, comas2008METMA}. These methods, however, are becoming increasingly important for the analysis of function-valued outcomes associated with spatial event-type data.  
Different from geostatistical or spatial areal data, the locations of the spatial point patterns themselves, at which we measure the (function-valued) objects, are treated as random. 

Although the developed methods allow for the analysis of spatial correlation and variation of the function-valued marks over space, they are, in general, not applicable for network-constrained data, where the locations of the function-valued quantities are governed by the configuration of a spatially embedded relational system. Indeed, despite the increasing availability and interest in identifying structures within such data sources, the analysis of network-constrained data with function-valued marks remains a highly daunting task and an extensively open area in contemporary urban data applications. 

Here, we focus on a particular challenge of identifying dependence and variation in bike-sharing system data in an urban context. In particular, using monthly data from a bike-sharing system as a data application, we extend recent results from marked spatial point processes to network-constrained data to investigate the structural interrelations between the monthly cycling profiles for a set of bicycle departure stations. In order to tackle this challenge, we frame the data problem as a point process constrained by a network, with marks represented by functions.

In the classic form, a marked spatial point process is characterised by a random collection of points distributed on a two-dimensional Euclidean space and usually a single scalar-valued mark, either an integer- or real-valued quantity providing further point-specific information \citep[see][for general treatmens]{Illian2008}. Different from the (Euclidean) spatial case,  adaptations of marked point process characteristics to point processes on linear networks remain elusive, and most contributions are restricted to unmarked settings \citep{moradi2018spatial, 10.1093/biomet/67.2.261}. Substituting the Euclidean distance by the shortest-path distance, \cite{OY01} and \cite{XZY08} were the first to propose a network-based version of Ripley's $K$-function and a kernel-based intensity estimator. Their works were later extended to geometrically-corrected first- and second-order summary characteristics, which inherently control for the relational structure of the network (see \citet{Ang2012, MFJ18}). More recent topics in this branch of research include the derivation of regular distance-based summary characteristics for linear networks \citep{rakshit2017second, cronie2020inhomogeneous}, the (fast) intensity estimations through kernel functions \citep{mcswiggan2017, rakshit2019fast},  Voronoi tessellations \citep{MoradiVor2019, MATEU2020Pseudo-separable}, penalised splines \citep{10.1214/21-AOAS1522}, directional analysis \citep{Moradidirectional}, directed linear networks \citep{DANGELO2021100534, Rasmussen2021}, spatio-temporal point patterns on linear networks \citep{MoradiMateu2020}, and a recent cross-validation-based statistical theory \citep{cronie2023cross}. In marked spatial point processes on linear networks, most contributions focus on integer-valued marks \citep{Spooner2004, BaddeleyJammalamadakaNair2014, EckardtJCGS, Eckardt2020}, whereas extensions to real-valued marks are recently proposed by \cite{Eckardt:Moradi:currrent, Eckardt2024Rejoinder}.  With the aim of extending the results of \cite{Eckardt:Moradi:currrent} to the settings where marks are function-valued quantities  \cite{Eckardt2023MultiFunctionMarks}, in this paper, we introduce various novel mark summary characteristics for point processes on linear networks with function-valued marks, widening the toolkit for the analysis of marked spatial point processes when the points live on linear networks, and the marks have some functional support.

Section 2 first presents some background material for marked point processes and then the corresponding characteristics for function-valued marks. A simulation study is shown in Section 3, while an application to real data comes in Section 4. The paper ends with some final discussion. 

\section{Methodology}\label{sec:methods}

\subsection{Preliminaries}\label{sec:Prelimlpp}

Let $\LL \subset \R^2$ be a linear network, and $X=\{ x_i \} = \lbrace (x_i, h(x_i)) \rbrace_{i=1}^N, 1\leq N < \infty$ denote a marked spatial point process on $\LL \times \F(\T),~\mathcal{T}=(a,b), -\infty\leq a\leq b\leq \infty$ with the points $x_i$ and the associated function-valued marks $h(x_i): \mathcal{T}\subseteq \mathds{R} \mapsto \mathds{R}$. In general, we assume $\LL$ to be equipped with a regular distance metric $d$, e.g. the shortest-path distance \citep{rakshit2017second}. Formally, $\LL=\bigcup_{j=1}^k l_j$ is defined as a finite union of some non-intersecting line segments $l_j=[u_j,v_j]=\{tu_j + (1-t)v_j:0\leq t\leq 1\}$, $u_j\neq v_j\in\R^2$ with total length $|\LL|=\sum_{j=1}^k |l_j|$ where $|l_j|=d(u_j,v_j)$. Further, $\F$ is assumed to be a complete separable metric space equipped with the $\sigma$-algebra $\mathcal{F}$ \citep{Daley2008}. 
%\jorge{here we use d() for the distance over $\LL$, should we use $d_\LL$? If so, later in eq (7) we use $d_L$ and it should be $d_\LL$. Also in the uni-dimensional kernel}
%Given $X$, let $\x$ and $\X$ denote an observed point pattern on $\LL$ and the ground process of $X$, i.e. its unmarked version, respectively, and use  $\lambda(\cdot)$ to denote the intensity function of $\X$. 
Fixing $F\in\mathcal{F}$, the expected number of points $N(\cdot)$, with function-valued marks in $F\in \mathcal{F}$, falling  in $A \subset \LL \times \F(\T)$ is
%per unit length of the network in the vicinity of a location $u \in \LL$ is given by
\begin{eqnarray}\label{eq:markintense}
  \
  \ee\left[
  N(X \cap A)
  \right]
  % =
  % \Lambda(X \cap A)
  % =
  % \int_{A}
  % \lambda \left( u \right) \de_1 u P(\de F)
  =
  \int_{A}
  \lambda \left( u, h(u) \right) \de_1 u P(\de F),
\end{eqnarray}
where $P(\de F)$ is a reference measure on $(\F,\mathcal{F})$, $\lambda(\cdot)$ is the intensity function of $X$, and $\de_1$ stands for integration with respect to arc length on the network. Note that the above definition allows for straightforward generalisation to $d$-variate function-valued attributes on $\F^d$ with $\sigma$-algebra $\mathcal{F}^d=\bigotimes^d_{l=1}\mathcal{F}_l$. 
%Considering $\X$, \eqref{eq:markintense} simplifies to   
% \begin{eqnarray}\label{eq:lambdanet}
%     \ee
%     [N(\X \cap A)]
%     =
%     \int_A \lambda(u) \de_1 u,
%     \quad
%     A \subset \LL,
% \end{eqnarray}
Denoting the $m$-order product intensity of $X$ by $\lambda^{(m)}(\cdot)$ and following   Campbell’s formulae, for any non-negative measurable function $f(\cdot)$ on the product space $(\LL \times \F(\T))^m$, we have
 \begin{align}\label{eq:product}
&
\ee
\left[
\mathop{\sum\nolimits\sp{\ne}}_{
(x_1, h(x_1)),
\ldots,
(x_m, h(x_m))
\in X}f
\left(
(x_1, h(x_1)),
\ldots,
(x_m, h(x_m)
\right)
\right] \nonumber
\\
&=
%\int_L\cdots\int_L 
\int_{(\LL \times \F(\T))^m}
f
\left(
(u_1, h(u_1)),
\ldots,
(u_m, h(u_m)
\right)
\lambda^{(m)} 
\left(
(u_1, h(u_1)),
\ldots,
(u_m, h(u_m)
\right)
\prod_{i=1}^m \de_i u_i P(\de F_i)
%\de_1u_1 \cdots \de_1u_m ,
\end{align}
and
\begin{align}\label{eq:correlnetwork}
g^{(m)}
\left(
(u_1, h(u_1)),
\ldots,
(u_m, h(u_m)
\right) 
   & =
    \frac{
    \lambda^{(m)} 
    \left(
(u_1, h(u_1)),
\ldots,
(u_m, h(u_m)
\right)
    }{
    \lambda(u_1, h(u_1)) \cdots \lambda(u_m, h(u_m))
    } \nonumber
    \\
    & =
\frac{
\lambda^{(m)}_{\LL} 
\left(
u_1, \ldots, u_m\right)
}{
\lambda_{\LL}(u_1) \cdots \lambda_{\LL}(u_m)
}
\frac{
r^{(m)} (h(u_1), \ldots, h(u_m) | u_1, \ldots, u_m)
}{
r(h(u_1)|u_1) \cdots r(h(u_m)|u_m)
} \nonumber
 \\
    & =
g^{(m)}_{\LL} (u_1, \ldots, u_m)
\gamma^{(m)} (h(u_1), \ldots, h(u_m) | u_1, \ldots, u_m),
\end{align}
where $r^{(m)}$ is the conditional density for marks, and $g^{(m)}_{\LL}$ is the $m$-th order correlation function of the ground process, i.e. unmarked point process.
%where $\sum_{D_1,\ldots,D_j}$ ranges over all partitions $\{D_1,\ldots,D_j\}$ of $\{1,\ldots,m\}$ into $j$ non-empty and disjoint sets, and $N(D_j)$ is the cardinality of the index set $D_j$ \citep{cronie2020inhomogeneous}. 
 
Finally, to extend planar summary characteristics to linear network settings, the notion of stationarity needs to be introduced. While the constrained nature of $\LL$ has, so far, prevented direct extensions of the concept of stationarity to linear networks \citep{baddeley2017stationary} due to the lack of a proper transformation mechanism, \cite{cronie2020inhomogeneous} introduced a notion of pseudo-stationarity for point processes on linear networks. In particular, $X$ is called $k$-th order intensity reweighted pseudo-stationary whenever the product intensities $\lambda^{(m)}(\cdot)$, $1\leq m\leq k$, exist, $\bar\lambda=\inf_{ (u,h(u)) \in \LL \times \F(\T)}\lambda(u,h(u))>0$, and for any $m\in\{2,\ldots,k\}$ the correlation function $g_m(\cdot)$ satisfies
\begin{align}
    \label{IRMPSstrong}
g^{(m)}
\left(
(u_1, h(u_1)),
\ldots,
(u_m, h(u_m)
\right)
=
\bar g^{(m)}_{\LL}(d(u,u_1),\ldots,d(u,u_m))
\bar \gamma^{(m)}(
h(u_1), \ldots, h(u_m) | d(u,u_1), \ldots, d(u,u_m)
),
\end{align}
%\jorge{HAS THIS $\bar g^{(m)}$ NEED TO HAVE AS SUBINDEX $L$, as in (3)???}
for any $u\in \LL$ and some functions
$\bar g^{(m)}_{\LL}:[0,\infty)^m\to[0,\infty)$ and $\bar \gamma^{(m)}:\F(\T)^m \to [0,\infty)$ \citep{cronie2020inhomogeneous, Ottmardis}. Moreover, $X$ is said to be intensity reweighted moment {pseudo-stationary} (IRMPS) whenever $k$-th order intensity reweighted pseudo-stationary holds for any order $k\geq 2$. If $X$ is a homogeneous $k$-th order intensity reweighted pseudo-stationary process, $X$ is said to be $k$-th order pseudo-stationary. Finally, a moment pseudo-stationary point process $X$ is considered (strongly) pseudo-stationary if its moments completely and uniquely characterise its distribution. 

\subsection{Summary characteristics for function-valued marks on linear networks}\label{sec:fct:methods}

Similar to \cite{Eckardt2023MultiFunctionMarks} and considering a pointwise specification, let $h_1=h(x)(t)$ and $h_2=h(y)(t)$ denote the marks for the pair of points $x,y \in X$ with interpoint distance $d(x,y)=r$ at time $t\in \mathcal{T}$. Further, let $h_1$ and $h_2$ be arguments of the test function $\tf_f:\F \times \F \rightarrow \R$, and write $c_{\tf_f}(r)(t)=\mathds{E}[\tf_f(h_1,h_2)]$ and $c_{\tf_f}(t)=c_{\tf_f}(\infty)(t)$.
%for the conditional expectation of the test functions at a shortest-path distance $r$, and when $r$ tends to infinity. 
We then define a pointwise $\tf_f$-correlation function $\kappa_{\tf_f}(r)(t)$ as 
\begin{eqnarray}\label{eq:tfcorrNet}
\kappa_{\tf_f}(r)(t)
=
\frac{
c_{\tf_f}(r)(t)
}{
c_{\tf_f}(t)
    }.
\end{eqnarray}
In general, $c_{\tf_f}(t)$ corresponds to the expectation of any given test function under the marks' independence assumption, and $c_{\tf_f}(r)(t)$ is assumed to coincide with $c_{\tf_f}(t)$ if the marks are not interrelated, for which case $\kappa_{\tf_f}(r, t)$ becomes one. We note that \eqref{eq:tfcorrNet} translates into a global version $\kappa_{\tf_f}(r)$ that allows for similar interpretations (in a $L_2$ sense) as for the classic mark summary characteristics, through the integration of   \eqref{eq:tfcorrNet} over $\mathcal{T}$, i.e.,  
\begin{equation}\label{eq:tfcorrNet:global}
\kappa_{\tf_f}(r)
=
% \int_{\mathcal{T}} \ee\left[\tf_f(h_1,h_2)\right]\de t 
% \eqaldot
\int_{\mathcal{T}} \kappa_{\tf_f}(r)(t)\de t.    
\end{equation}
The specific linear-network-based mark summary characteristics and their interpretation as such are completely determined by the explicit formulation of the test function $\tf_f(h_1, h_2)$. A summary of pointwise generalisations for the most commonly applied test functions in the literature, together with their notation, is given in Table   \ref{tab:testfuns:nets}.
 Let $\mu_h(t)$ be the 
 %pointwise mark 
 mean and $\sigma^2_h(t)$ be the variance of the marks at time $t$. 
\begin{sidewaystable}[!h]
\caption{Pointwise test functions for point processes on linear networks with function-valued marks. The average and variance of all marks at $t\in\mathcal{T}$ are denoted by $\mu_{h}(t)$ and $\sigma^2_{h}(t)$, and $\mu_{h}(r)(t)$ is the conditional mean of the marks for points with an interpoint distance $r$ at $t\in\mathcal{T}$.
%for points on a linear network where $\left[\cdot\right]$ denotes operations with respect to the network distance $d$
}

\begin{center}
\begin{tabular}{l | ccc}
\hline
Name and symbol of  & Name of $\tf_f$ & pointwise test function & pointwise normalising \\
mark characteristic                     &   & $\tf_f(h_1,h_2)$  & factor $c_{\tf_f}(t)$ \\
\hline
Stoyan's mark correlation function $\kappa_{hh}(r)(t)$ & $\tf_1$ & $ h_1h_2$ &  $\mu_{h}^2(t)$
 \\
\hline
Beisbart and Kerscher's   mark correlation function $\kappa_{hh}^{\mathrm{Bei}}(r)(t)$ &   $\tf_2$& $h_1 + h_2$ &  $2\mu_{h}(t)$
\\
\hline
$\mathbf{r}$-mark correlation function $\mathbf{r}_{h \bullet}(r)(t)$ &  $\tf_3$& $h_1$ &  $\mu_{h}(t)$
 \\
 \hline
$\mathbf{r}$-mark correlation function  $\mathbf{r}_{\bullet h}(r)(t)$ &  $\tf_4$& $ h_2 $ &  $\mu_{h}(t)$
\\
\hline
Mark variogram $\gamma_{hh}(r)(t)$ & $\tf_5$ & $0.5
    (h_1-h_2)^2 $ &  $\sigma^2_{h}(t)$
\\
\hline
Stoyan's covariance function $\cov_{hh}(r)(t)$ &  $\tf_6$ & $h_1h_2-(\mu_{h}(t))^2$ &  1
\\
\hline
Isham's   mark correlation function $\kappa^{\mathrm{Ish}}_{hh}(r)(t)$ &   $\tf_7$ & $h_1h_2$ $-(\mu_{h}(t))^2$&  $\sigma_{h}^2(t)$
\\
\hline
Schlather's $I$ function  $I_{hh}(r)(t)$ &   $\tf_8$& $(h_1 - \mu_{h}(r)(t)) (h_2 - \mu_{h}(r)(t))$ &  $\sigma^2_{h}(t)$
\\
\hline

Shimanti's $I$ function   $I^{\mathrm{Shi}}_{hh}(r)(t)$ &  $\tf_{9}$ & $(h_1 - \mu_{h}(t)) (h_2 - \mu_{h}(t))$ &  $\sigma^2_{h}(t)$
\\
\hline
\end{tabular}
\label{tab:testfuns:nets}
\end{center}
\end{sidewaystable}

Specifying $\tf_f=\tf_1$, computation of \eqref{eq:tfcorrNet:global} yields a network-based version of Stoyan's mark correlation function $\kappa_{hh}(r)$, which, taking the geometry of linear networks into account and employing a regular metric \citep{rakshit2017second}, helps to investigate the average pairwise association of the function-valued marks of the points on the network. If the function-valued marks are independent, their product tends to  $\mu_h^2=\int_{\mathcal{T}}\mu_h^2(t)\de t$ such that $\kappa_{hh}(r)$ is equal to one. Applications of the test function $\tf_2$ lead to
%a computational simpler version of Stoyan's mark correlation function $\kappa_{hh}(r)$ 
the
Beisbart and Kerscher's mark correlation function \citep{Beisbart:2000} where  $\kappa_{hh}^{\mathrm{Bei}}(r)$ is constantly equal to one under mark independence. 
Both $\tf_3$ and $\tf_4$ relate to the expected value of either $h_1$ or $h_2$ with respect to the shortest-path distance $r$ and are usually different from the pointwise mean $\mu_h(t)$ except under the assumption that marks are independent. The mark variogram $\gamma_{hh}(r)$, which, instead of the association, quantifies the variability of the function-valued marks for pairs of points with an interpoint distance $r$, is obtained by specifying $\tf_f=\tf_5$ and computing \eqref{eq:tfcorrNet:global}. If the function-valued marks are similar for any pair of points with an interpoint distance $r$,  their dispersion is small. For independent marks, the pointwise half-squared distance between the function-valued marks tends to the variance $\sigma^2_h(t)$, and $\gamma_{hh}(r)$ is equal to one. A similar quantity, Stoyan's covariance function $\cov_{hh}(r)$ results from choosing $\tf_6$ and extends to Isham's   mark correlation function $\kappa^{\mathrm{Ish}}_{hh}(r)$ by normalising $\cov_{hh}(r)(t)$ by $\sigma^2_h(t)$. Finally, the selection of $\tf_8$ and $\tf_9$ in \eqref{eq:tfcorrNet} yields point process versions of Moran's $I$ \citep{moran}. 
%Although similar in spirit, $\tf_8$  and $\tf_9$ apply a different centring of the marks.
While  $\tf_9$ centres $h_1$ and $h_2$ by $\mu_h(t)$, $\tf_8$ uses the pointwise conditional mean $\mu_h(r)(t)$ which is the mean of function-valued marks of all points with an interpoint distance $r$ at time $t$. 

%\matthias{We note that, alternatively, the data could also be treated as a high-dimensional spatio-temporal point process $\lbrace x_i(s), m(x_i(s))\rbrace$ with $i = 1,\ldots, N$ points and associated real-valued marks $m(x_i(s))$ recorded at $s=1,\ldots, \mathcal{S}\subseteq \R$ distinct time stamps, where $N<\mathcal{S}$ on $\R^2\times \M\times \R$. However, this is computationally burdensome and might become even NP-hard.} 

\subsection{Estimation}

For distinct  points $(x,h(x)(t)), (y,h(y)(t)) \in X$, $t\in\mathcal{T}$, and a given test function $\tf_f$, all mark summary characteristics presented in Table \ref{tab:testfuns:nets} can be estimated through the $\tf_f$-correlation function in \eqref{eq:tfcorrNet}, as
\[
\hat{\kappa}_{\tf_f}(r)(t)=\frac{\hat{c}_{\tf_f}(r)(t)}{\hat{c}_{\tf_f}(t)},
\]
where %$\hat{c}_{\tf_f}(r)(t)$ takes the form 
\begin{align}
\hat{c}_{\tf_f}(r)(t) =    \frac{
    \sum_{(x, h(x)(t)),(y,h(y)(t))\in X}^{\ne}
\tf_f(h(x)(t),h(y)(t))\mathfrak{K}(d(x,y)-r)
    }{
    \sum_{(x, h(x)(t)),(y,h(y)(t))\in X}^{\ne}
\mathfrak{K}(d(x,y)-r),
    }
\end{align} 
%\jorge{Check if we have to use $d_\LL$, and $\mathfrak{K}_\LL$ }
and
\begin{align}
    \hat{c}_{\tf_f}(t) 
    = 
    \frac{
    \sum_{(x, h(x)(t)),(y,h(y)(t))\in X}^{\ne} \tf_f(h(x)(t),h(y)(t))
    }{
    N^2
    }
\end{align}
where $\mathfrak{K}$ is a uni-dimensional kernel function \citep{Baddeley2010}. Again, the specific mark characteristics are determined by the specification of the test function $\tf_f(h(x)(t),h(y)(t))$.
%for the marks $h(x)$ and $h(y)$ at the shortest path distance $\Vert x-y \Vert = r$. 
Analogous to Section \ref{sec:fct:methods}, the pointwise estimators can then be translated into global mark summary characteristics through  
\[
\hat{\kappa}_{\tf_f}(r)=\int_a^b \hat{\kappa}_{\tf_f}(r)(t)\de t.
\]

\section{Simulation study}\label{sec:simu}
To evaluate the performance of the proposed summary characteristics under different settings, we consider three scenarios. Throughout the simulation studies, we make use of the street network of Vancouver, Canada, which is used in our real data analysis; details of the network are given in Section \ref{sec:realdata}. We generate a realisation of a homogeneous Poisson process with constant intensity function $\lambda=0.0006$ leading to 356 points, for which we consider three different scenarios for function-valued marks, {all respecting the dimension of 30 timestamps}.
\begin{itemize}
    \item {Scenario} One: marks are independently generated from a continuous uniform distribution $U(0,1)$.
   
    \item {Scenario} Two: The marks in scenario one are multiplied by the spatial distance between the points and the border of the network represented by nodes with only one outgoing segment.
    
    \item {Scenario} Three: For each point, let $\alpha$ be the number of neighbouring points within a spatial distance of 876 units. A continuous uniform distribution with parameters $\alpha-\alpha/2$ and $\alpha+\alpha/2$ is used to generate the function-valued marks.
\end{itemize}

\begin{figure}[!h]
\centering
\includegraphics[scale=.115]{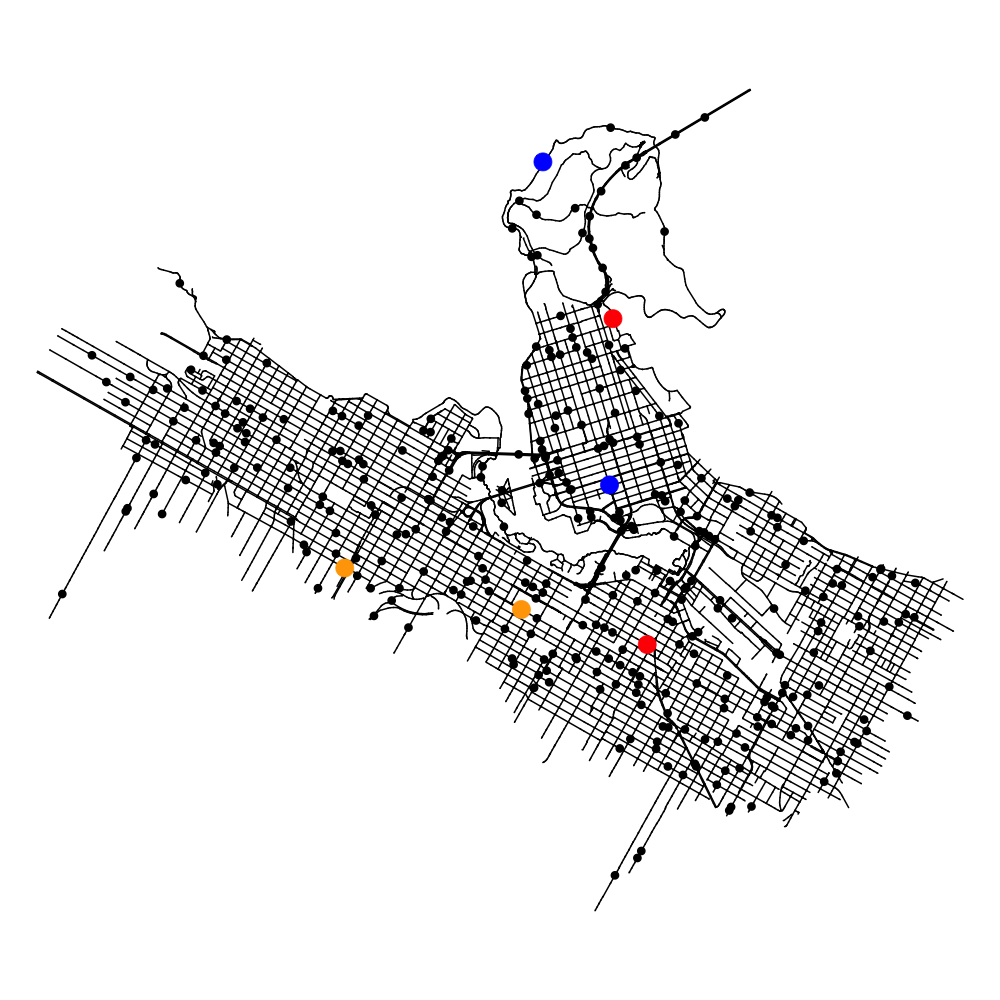}
\qquad
\includegraphics[scale=.115]{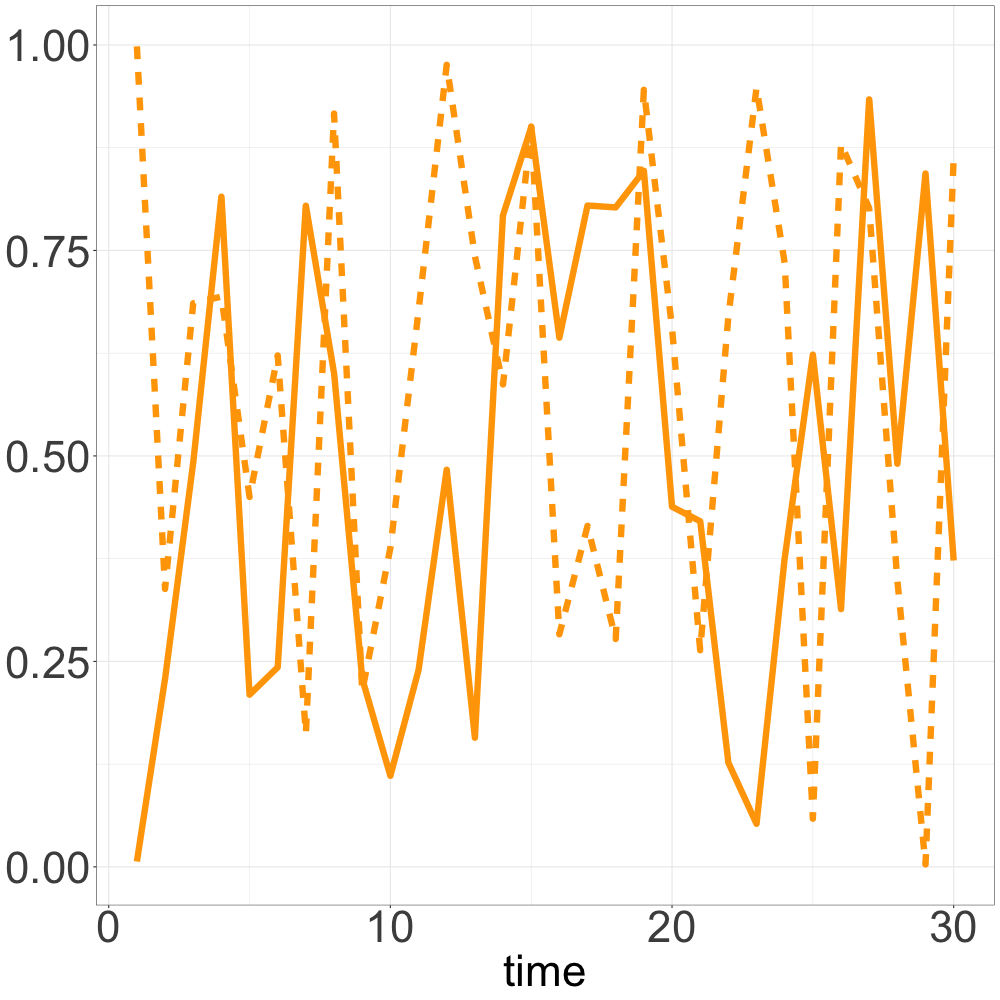}
\\
\includegraphics[scale=.115]{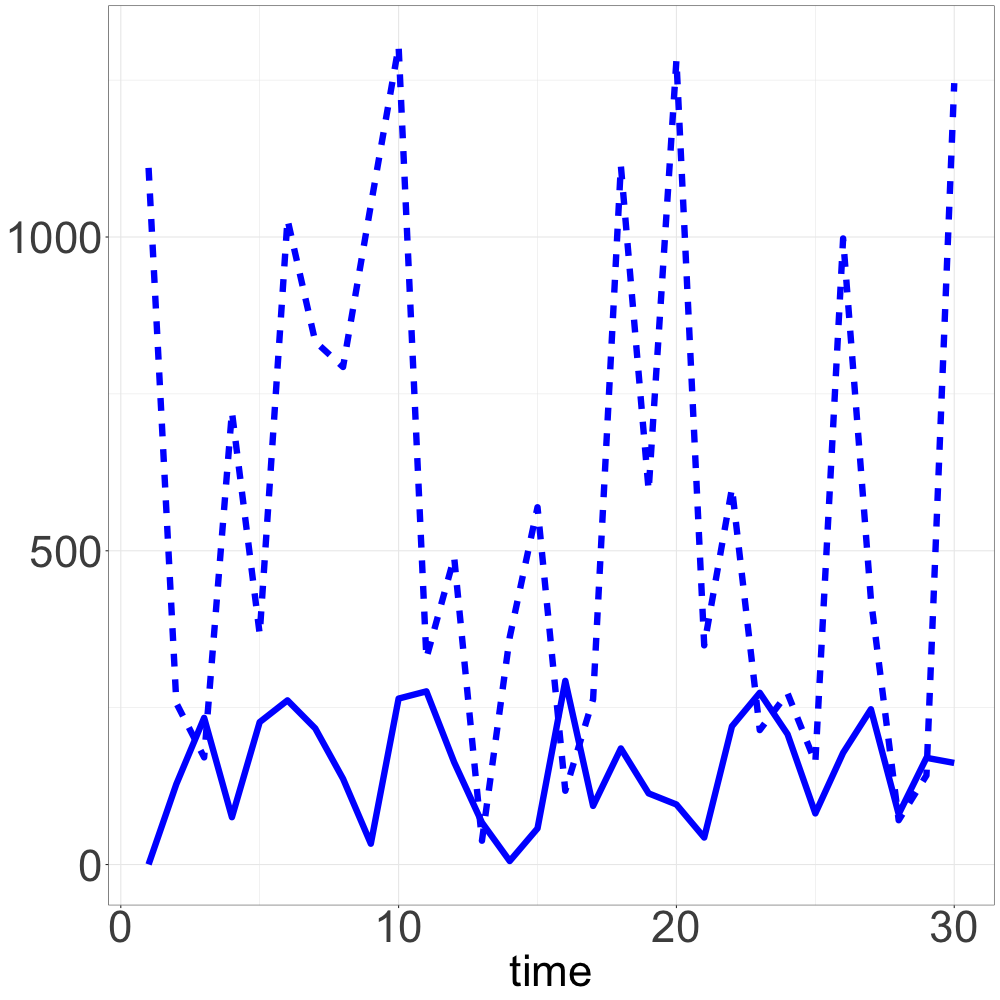}
\qquad
\includegraphics[scale=.115]{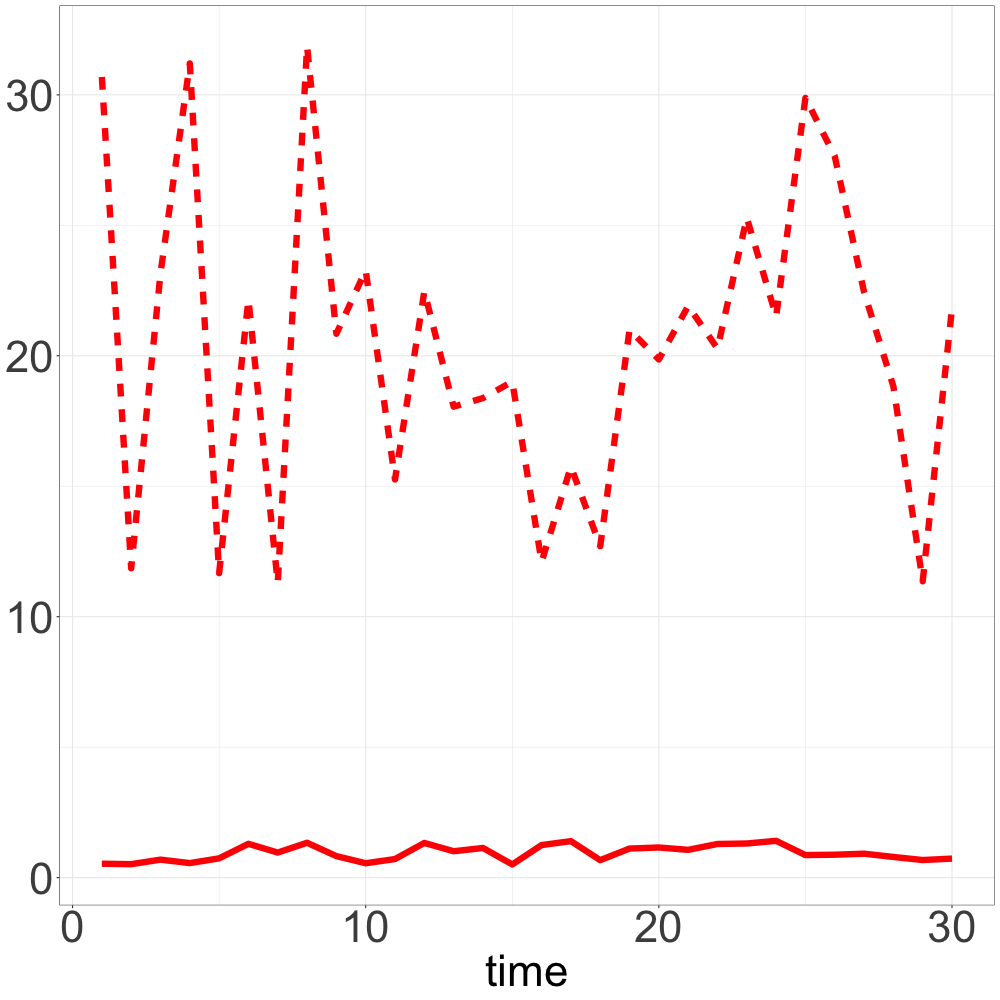}
\caption{
The simulated point pattern and two arbitrary points per scenario to see the differences between Scenario One (orange), Scenario Two (blue), and Scenario Three (red).
}
\label{fig:datasim}
\end{figure}

Figure \ref{fig:datasim} showcases the simulated point pattern alongside functional marks for two arbitrary points per scenario,  highlighting the existing differences in spatial structure among the three scenarios. In Scenario One, marks exhibit no discernible spatial pattern, while in Scenario Two, points with large marks are observed farther from the network's border, and in Scenario Three, marks tend to be larger for points with more neighbours and smaller for those with less neighbours.

\begin{figure}
\centering
\includegraphics[scale=.75]{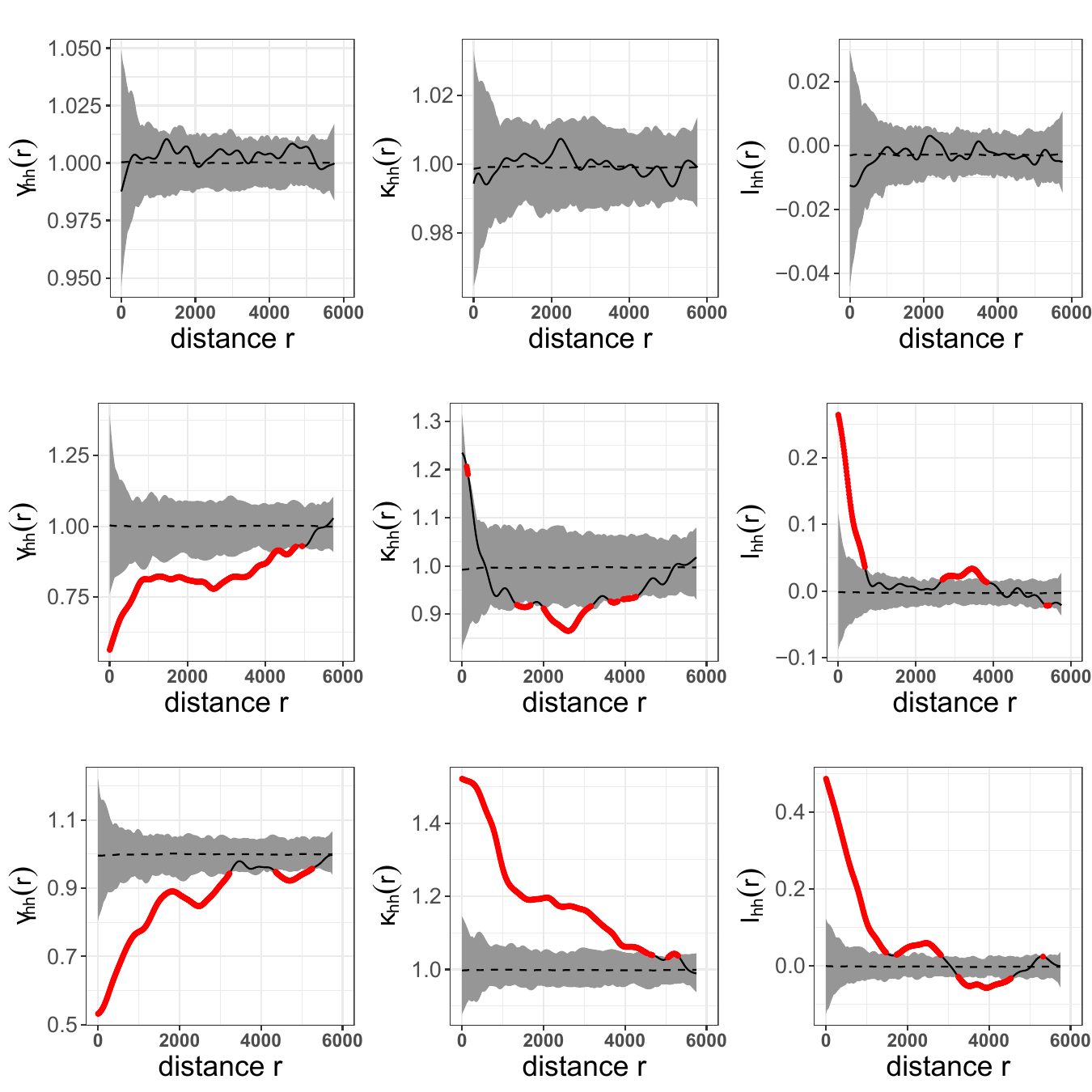}
\caption{
Normalised mark summary characteristics with $95\%$ global envelopes using 500 permutations computed from 356 simulated data points of a homogeneous Poisson process on the Vancouver traffic network with independent and identically distributed (\textit{top row}) and dependent (\textit{central and bottom rows}) function-valued marks: Mark variogram (\textit{left column}), Stoyan's mark correlation function (\textit{central column}), and Shimantani's $I(r)$ function (\textit{right column}).  
\label{fig:sims}
}
\end{figure}

The simulated marked point patterns are then used as input of the mark variogram $\gamma_{hh}(r)(t)$, the mark correlation function $\kappa_{hh}(r)(t)$ and the adapted version of Shimantani's $I_{hh}(r)(t)$ function. The results with 95\% global envelopes based on the extreme rank length (ERL) measure \citep{mari1, MrkvickaEtal2020} using 500 permutations of the function-valued marks are depicted in Figure \ref{fig:sims}.  
At first, one can see that in the case of the independent mark scenario (top panels), all three empirical summary characteristics fall within the global envelopes, which represent random labelling, indicating that there is no particular association or spatial structure between the function-valued marks in Scenario One. In contrast, all three plots of the central and the bottom panels show clear deviations from random labelling, represented by the $95\%$ global envelopes, indicating that there are {significative spatial associations amongst the} function-valued marks. 
More specifically, in scenario two (central panel), looking at the mark variogram, we can see that for small distances, it falls below the global envelop highlighting that points with a small interpoint distance have similar marks, and consequently, their corresponding half-squared distances are smaller than those expected under of random labelling; this is in agreement with how marks are generated as nearby points have a similar travelling distance to the border of the network. Turning to the mark correlation function, we can see that the marks for points at rather small distances are, on average, larger than under the case of random labelling; the number of points located far from the border of the network is more than the number of points locating near the border. Increasing the interpoint distance will result in more points contributing to the mark correlation function, for which at least one has a small mark. The same is indicated by the proposed version of Shimantani's $I_{hh}(r)(t)$.
In the case of scenario three (bottom panel), we can see that the behaviour of the mark variogram is similar to that of scenario two, meaning that the number of neighbours for points with small interpoint distances is similar. The mark correlation function falls above the global envelope, showing that for two points with small to moderate interpoint distances, at least one of them has a large number of neighbours. This behaviour is also indicated by the proposed version of Shimantani's $I_{hh}(r)(t)$. In general, from the three considered scenarios, one can see that the association among points is better represented based on small to moderate interpoint distances.

%show very small envelopes and support the independent mark assumption as simulated in the data generation. 
 
%  \textcolor{blue}{For nearby points, the average half-squared distances are smaller than expected under the independent mark assumption as indicated by the mark variogram (left). Likewise, the mark correlation function for scenario 2 where the mark is constructed from the distance to the boundary of the network, reveals that the marks for points at rather small distances are on average larger then expected. With increasing distances, when the distance to the boundary begins to vary more strongly, the average product of the marks shows a negative clear deviation from the global 95\% envelopes. Both Shimantani's $I$  functions (central and bottom right panels) highlight a strong but decreasing positive spatial autocorrelation for the marks at smaller distances.}     

% \jorge{We did not mention scenarios 2 and 3 for the dependent cases. We only write: ...we additionally weighted the constructed marks by the minimal distance of the station to the borders of the Vancouver network...I guess the central row of Fig 1 corresponds to this scenario, and the bottom row to what you write as scenario 3? Please rewrite and clarify }

% Scenario 3 is computed by first summing the number $N_P$ of points locations that exceed the distance of $r = 876$ for each point. In a subsequent action, we sampled from a uniform distribution with parameters $a=N_p-N_p/2$ and   $b=N_p+N_p/2$.  

\section{Vancouver cycle distance profiles 
}\label{sec:realdata}

\begin{sidewaysfigure}
\centering
\includegraphics[scale=.74]{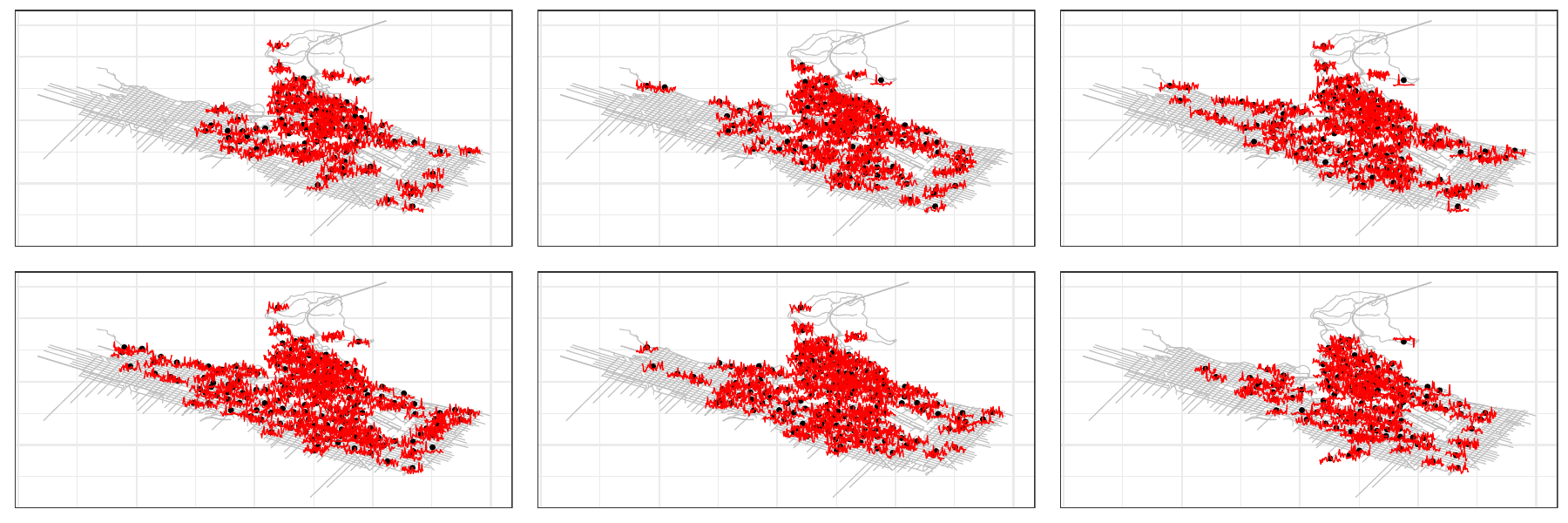}
\caption{
Locations of departure stations on Vancouver's street network with bicycle distance profiles as function-valued marks from May (top-left) to October (bottom-right) 2022. 
%: \jorge{Patterns from May to July (\textit{top}) and from August to October (\textit{bottom}), and from \textit{left} to \textit{right}}
}
\label{fig:profiles}
\end{sidewaysfigure}

For an illustration of the proposed methods, we consider monthly bike system data collected by Vancouver's public bike share program {\em Mobi by Shaw Go} provided under a public data license agreement \footnote{\url{https://www.mobibikes.ca/en/system-data}}. This data has previously been studied through non-spatial statistical methods to identify the characteristics of  \textit{super users} \citep{Winter}, cycling behaviour \citep{HostfordEval} and the spatial access to bike sharing system \citep{HosfordWinter}. Initiated in 2006, the bike-sharing system involved a network of 250 docking stations distributed across the city, located approximately every second or third block or at an inter-station distance of around 200 to 300 meters. For each trip, the provided data reports the locations of docking stations where a trip is started/ended, the exact time stamps, the covered cycled distance in meters, the trip duration in seconds, the battery voltage at departure and return, the temperature at both docking locations, and further stop specific information. Restricting the analysis of system data recorded from May to October 2022, our initial sample considered the information for 735,739 distinct bicycle trips. Separated by month, we found the difference in the average cycled distance per station among months is minor, while both the number of bicycle trips and the number of stations show a clear heterogeneity between the months; see Table \ref{tab:summary:dats} for a summary of the data under study. 
\begin{table}[!h]
\caption{Summary of the bike system monthly data
}
\begin{center}
\begin{tabular}{lccc}
\hline
Month & Number of point locations & Number of trips & Average distance (m) \\
\hline
May & 121 & 76,645 &  2,795 \\
June & 147 & 105,055& 2,858\\
July & 159 & 144,122 & 2,991 \\
August & 191 & 164,502& 2,974\\
September & 186 & 134,959 & 2,788 \\
October & 136 & 110,456 &  2,704\\
\hline
\hline
\end{tabular}
\label{tab:summary:dats}
\end{center}
\end{table}
%Next, we selected the exact address of each departure location and linked the obtained address information to the corresponding spatial coordinates of the departure stations.
In order to build function-valued marks for each station, we calculate the average cycling distance per day, which leads to monthly cycling profiles per station. Except for some extreme peaks, the station-specific distance profiles behave similarly over the six months under study; see Figure \ref{fig:profiles}.

%Finally, we aggregated the covered distance per day to obtain the station-based average cycling distance curves per month and assigned the monthly distance profiles as function-valued marks to point locations. Except for some extreme peaks, the station-specific distance profiles behave similarly over the six months under study.   
\begin{figure}
\centering
\includegraphics[scale=.4]{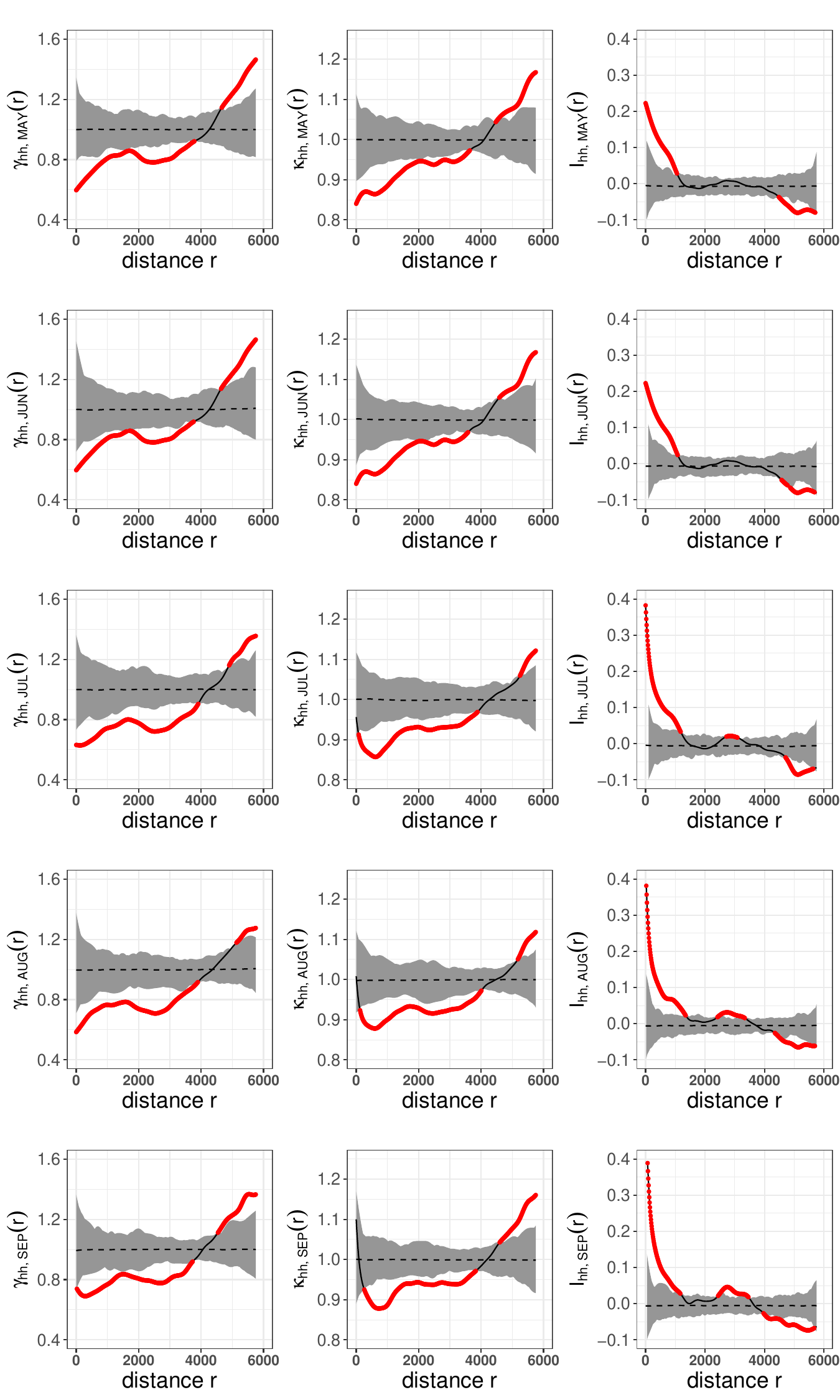}
\caption{
Normalised mark summary characteristics with $95\%$ global envelopes using 500 permutations computed from the Vancouver bike-sharing data from May to October 2022 with the averaged daily cycling distances as function-valued marks. Mark variogram $\gamma_{hh}(r)$ (\textit{left}), Stoyan's mark correlation function $\kappa_{hh}(r)$ (\textit{middle}), and Shimantani's $I_{hh}(r)$ function (\textit{rigth}) from May (\textit{top}) to October (\textit{bottom}) 2022.
}
\label{fig:Results}
\end{figure}

%\jorge{IT WOULD BE NICE TO HAVE ONE FIG ZOMMING IN A REGION AND SEE BETTER THE FUNCTIONAL MARKS}. 
Using the created function-valued marked point patterns as input, we now compute the mark variogram $\gamma_{hh}(r)$, the counterpart version of Stoyan's mark correlation function $\kappa_{hh}(r)$, and Shimatani's $I_{hh}(r)$ per month. For each month, we additionally construct $95\%$ global envelopes using 500 permutations of the function-valued marks and the R \citep{Rcore} package \texttt{GET} \citep{GETpack} to test for the random labelling hypothesis; results are given in Figure \ref{fig:Results}. All three characteristics highlight clear deviations from the random labelling assumption, together with some variation from May (left) to October (right) 2022. The mark variogram (top panels) shows the average variation of the distance profiles for any pair of distinct points as a function of the network distance $r$; all six plots suggest less variation among distance curves compared to the expected one under the random labelling assumption up to $r = 4000$. In particular, for smaller distances up to $r = 1000$, the distance profiles appear to be less heterogeneous. This result might be explained by short-distance trips within the city centre of Vancouver, which has a high frequency of nearby locking stations. In contrast, with increasing distance $r\geq 5000$, all six plots show a clear positive deviation from the global envelopes, highlighting a strong heterogeneity between the distance profiles for any pair of bike stations which have an interpoint distance larger than $r\geq 5000$ meters.    
A similar tendency among the distance profile is suggested by Stoyan's mark correlation functions depicted in the central panels of Figure \ref{fig:Results}. For distances up to $r=4000$, all mark correlation functions suggest that the pairwise product of the function-valued marks is smaller than expected under the random labelling assumption. This would correspond to small distance profiles for nearby locking stations. Again, an opposite tendency is indicated for distances $r\geq 5000$ where all summary characteristics show a clear positive deviation from the global envelopes. Finally, the adapted version of Shimantani's $I$ function (right panels) reveals clear positive and also negative deviations from the random labelling assumption for distances $r\leq 1000$ and $r\geq 4500$. Recalling the close relation of Shimantani's $I$ to Moran's autocorrelation index, these results would correspond to a high positive spatial autocorrelation between nearby stations contrasted with negative autocorrelation between pairs of stations with a large interpoint distance. Note that large-distance bike trips usually correspond to travelling from downtown Vancouver to a destination outside the central area.
%While nearby inter-station distances are placed with the city centre, pairs of faraway locking stations would relate locking stations within the central city centre with stations at the city borders with corresponding homogeneous and heterogeneous cycling distance profiles, respectively. 
Comparing all three mark characteristics over the six months under study, we observe that cycling profiles in May and June exhibit similarities, with slight variations compared to July, August, September, and October.

%most variation in the empirical curves is revealed between May and June on the one hand to July, August and September on the other hand. 

\section{Discussion}\label{sec:discuss}

This paper introduces a range of mark summary characteristics tailored for network-constrained point processes with function-valued marks, 
offering valuable insights into the spatial variation and the correlations amongst marks. Furthermore, they serve as effective tools for discerning spatial dependencies and similarities among the observed curves, with potential applications across diverse fields due to their general applicability. Recall that in our setting, marks refer to functions of time.

Here, we have only focused on scenarios where a single function-valued quantity is observed at each point location. Through a simulation study, where we considered different settings and a real data analysis focused on cycling distance profiles within the city of Vancouver, Canada, we observed that our proposed summary characteristics adequately disclose the association among function-valued marks. In particular, mark variogram $\gamma_{hh}(r)$, Stoyan's mark correlation function $\kappa_{hh}(r)$, and Shimantani's $I_{hh}(r)$ function were used, and all of them were successful in differing any structure between the function-valued marks from random labelling.
As referred to here, we can build a large number of summary characteristics, although to be focused, we have only used three of them in practice. In this line, this paper delineates the path to follow in case other characteristics are needed, and thus, we have enlarged the current possibilities of working with (functional) marked point patterns.

As for future works, it would be interesting to expand the proposed mark summary characteristics to situations where each point is labelled by multiple function-valued marks or combinations of diverse object-valued marks. These could include scalars, functions, or compositions, with the potential to enrich our understanding of spatial data analysis when points are labelled by marks of different types.

\section*{Acknowledgments}

The authors gratefully acknowledge financial support through the German Research Association. Matthias Eckardt was funded by the Walter Benjamin grant 467634837 from the German Research Foundation. Jorge Mateu was partially funded by the Ministry of Science and Innovation (PID2022-141555OB-I00), and Generalitat Valenciana (CIAICO/2022/191). 

\newpage
\bibliographystyle{ecta}
\bibliography{FMLpp}
\end{document}